\documentstyle[aps,prl,floats,twocolumn,epsf]{revtex}
\def\eq{\begin{equation}}
\def\en{\end{equation}\noindent}
\def\eqa{\begin{eqnarray}}
\def\ena{\end{eqnarray}}

\def\LR{({L \atop R}) }
\begin{document}
\draft

\twocolumn[\hsize\textwidth\columnwidth\hsize\csname@twocolumnfalse%
\endcsname

\title{\bf A Ball in a Groove}
\author{Thomas C. Halsey and Deniz Erta{\c s}}
\address{Exxon Research and Engineering, Route 22 East, Annandale,
N.J. 08801}
\date{\today}
\maketitle

\begin{abstract}
We study the static equilibrium of an elastic sphere held in a rigid groove by
gravity and frictional contacts, as determined by contact mechanics. As a function of the
opening angle of the groove and the tilt of the groove with respect to the
vertical, we identify two regimes of static equilibrium for the ball. In the
first of these, at large opening angle or low tilt, the ball rolls at both
contacts as it is loaded. This is an analog of the ``elastic" regime in the
mechanics of granular media. At smaller opening angles or larger tilts, the ball
rolls at one contact and slides at the other as it is loaded, analogously with
the ``plastic" regime in the mechanics of granular media. In the elastic
regime, the stress indeterminacy is resolved by the underlying kinetics of the
ball response to loading.
\end{abstract}
]


The behavior of granular materials is one of the most difficult subjects in
solid mechanics \cite{general}. It is elementary to state the problem--given a
packing of elastic bodies, whose contacts with one another obey the classical
Amonton's laws of friction, what is the nature of the stress transmission in the
system in the limit where the forces between the bodies are much smaller than the
appropriately dimensioned elastic moduli of the bodies? This
problem underlies much of soil mechanics, sedimentology, and powder
technology, but has thus far resisted {\it a priori} solution, although much is
known at an empirical level about the behavior of such systems \cite{nedderman}.

The difficulty of this problem arises from its non-linearity, which appears in
two different guises. In the first place, the forces between elastic bodies in
contact are non-linear functions of their displacements. In the
second case, sufficiently strong shear forces can cause a contact to fail,
which is also a source of non-linearity.

Soil mechanicists often focus on the second source of non-linearity, but
presume that at a macroscopic level the first source can be finessed. The
result is a class of ``elasto-plastic" models that focus on the interplay
between an assumed linear elastic region of soil response and the plasticity
connected with shear-induced failure \cite{nedderman,elastoplastic}. While this
is a plausible and effective approach, it is disturbing to realize that there is
no fundamental reason to expect linear elastic behavior in a granular system at
all, given the non-linearity of contact mechanics. This uncertainty in the
fundamental underpinnings of granular statics is sometimes termed the ``stress
indeterminacy" problem, as the stress tensor cannot be determined from the
equations of force balance alone; these must be supplemented with some
constitutive relations \cite{cates}.

This uncertainty has motivated statistical
approaches to the problem, notably the ``$q$ -model" and its variants, which focus on
the fluctuations of forces within a granular packing \cite{qmodel}. Unfortunately, these models
are not based on a realistic picture of the contact forces between particles.  

\begin{figure}
\centerline{\epsfxsize=3.1in \epsffile{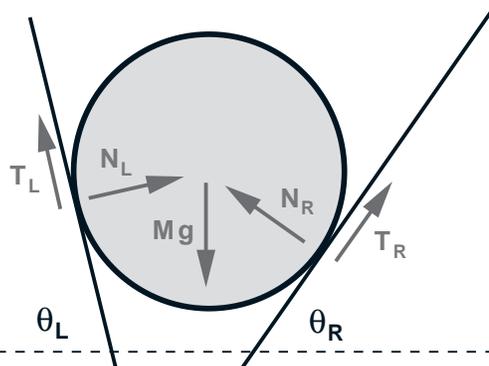}}
\caption{A ball in a groove is subject to five forces--the gravitational force,
which acts on its center of mass, and tangential and normal forces at each
contact. The two rigid planes are inclined at angles $\theta_{\LR}$ to the
horizontal.}
\label{ballngroove}
\end{figure}

For this reason, we here address a simple model system, which
nevertheless exhibits stress indeterminacy: a ball placed in a groove with
perfectly rigid walls (Figure
\ref{ballngroove}) under gravitational loading. Since the contact mechanics of
this system are tractable, we are able to solve its stress indeterminacy
problem. We find that it is possible to specify an ``elastic" and a ``plastic"
regime in the behavior of this ball, although the elastic regime is not closely
related to ordinary continuum elasticity. In the elastic regime, the stress
indeterminacy is lifted by the underlying kinetics of the ball as it is
loaded--since the number of kinematic parameters in this case is identical to
the number of constraints on the ball, we term this the {\it isokinetic} regime,
in contrast to the {\it isostatic} plastic regime
\cite{mouzarkel}, in which the number of independent forces equals the number of
constraints.

Suppose that, as in Fig.~\ref{ballngroove}, we have an elastic sphere of radius $R$, Young's
modulus $E$, Poisson ratio $\nu$, and mass $M$ in contact with a rigid groove.
We consider only motions in the plane perpendicular to the groove axis (out of the page in
Fig.~\ref{ballngroove}), thus reducing our problem to an essentially two dimensional one (note however, that
we still consider an elastic sphere--the contact mechanics of a disk are
essentially different.) $\theta_L$ and $\theta_R$ denote the angle between 
the left and right walls of the groove with respect to the horizontal, 
respectively. Without loss of generality we choose $\theta_L \ge \theta_R$. It is
convenient to define
$\gamma=
\frac{1}{2}(\theta_L +
\theta_R)$ and $\delta = \frac{1}{2}(\theta_L-\theta_R)$, so that $\pi-2\gamma$ is the
groove opening  angle, and $\delta$ gives the tilt of the groove with respect to
the vertical. Our definition of a groove satisfies $0 \le \delta \le \gamma \le
\frac{\pi}{2}$.  The normal and tangential forces exerted by
the wall upon the ball are $N_L$, $T_L$ on the left, and $N_R$, $T_R$ on the
right, with the tangential forces defined so that a force exerted upwards is
positive. The ball is also subject to the gravitational force $Mg$. The
static equilibrium of the ball requires that

\eqa
T_L \sin \theta_L + N_L \cos \theta_L + T_R \sin \theta_R + N_R \cos \theta_R
= Mg, \label{eq:equil1}\\
-T_L \cos \theta_L + N_L \sin \theta_L + T_R \cos \theta_R - N_R \sin \theta_R 
=
0, \label{eq:equil2}\\
T_L = T_R,
\label{eq:equil3}
\ena
corresponding to the requirements that the net horizontal and vertical forces,
and the net torque, be zero. In addition, Amonton's law requires that

\eq
T_{\LR} \le k N_{\LR},
\label{eq:Amont}
\en
where $k$ is the coefficient of friction between the ball and the wall. Since
we have three constraints for the four forces $T_L$, $N_L$, $T_R$, and $N_R$,
even subject to the Amonton's law inequalities we will have in general a
manifold of solutions. In order to determine which solution on this manifold is
actually chosen we must examine in more detail the mechanics of the ball itself.


The problem of an elastic sphere compressed against a rigid flat surface was
given its classical solution by Hertz \cite{johnson}. 
If the sphere (of radius $R$) is loaded by a force normal to the wall $N$, then
its center will displace towards the wall by a distance $u$, forming thereby a
contact circle of radius $a$, with

\eqa
a = (GNR)^{\frac{1}{3}}, \\
u = \left ( \frac{GN}{\sqrt{R}} \right )^{\frac{2}{3}},
\label{eq:hertz}
\ena
where $G = 3(1- \nu^2)/(4E)$.

The situation in the presence of tangential forces is considerably more
complicated \cite{johnson}. If an already established normal contact is sheared
at a tangential force that is insufficient to cause failure, a region of
microslip adjacent to the outer perimeter of the contact forms, moving inwards as the
tangential force increases. In this region, the two surfaces slip with respect to one another.
In the interior of the contact, the two surfaces remain stuck together as the
tangential force is applied. Eventually, at failure, there is no stick
region left.

The details of this process were explored in a classic paper by Mindlin and
Deresiewicz \cite{md}. They observed that
the analysis of the contact structure at finite tangential and
normal force is only tractable in the limit of low Poisson ratio $\nu$. In this
limit the local normal and tangential strains at the contact decouple from one
another, so that the normal pressures at the contact are always
those of a Hertzian contact with the same overall normal force. Thus the
additional tangential strains do not alter the solution of the part of the
problem corresponding to the normal force.

The problem is essentially hysteretic, in that the strains
corresponding to a given tangential and normal force are determined by the
specific history of contact loading leading to these forces. A general
increment of force in such a loading history can be described by the
change in tangential force $\delta T$ and the change in normal force
$\delta N$. A special class of such increments do not involve microslip
over some area of the contact surface, those with \cite{md}

\eqa
\delta N &>& 0, \\
\vert \delta T \vert &<& k \delta N.
\ena

This latter result suggests that we specify a particularly simple loading history
of a contact, in which $T/N$ is maintained constant, with an absolute
value $\le k$, while one force, say $N$, is monotonically (and adiabatically)
increased as a function of time. This type of loading history we term {\it
proportional loading}.

For proportional loading, we can also specify the compliance of the contact, or
the values of $u_n$, $u_t$, which are respectively the normal and tangential
displacements of the center of the ball corresponding to fixed forces $N$ and
$T$. From the Mindlin-Deresiewicz results, it is straightforward to show that
for small Poisson ratio,

\eq
\frac{u_t}{u_n} = \frac{T}{N},
\label{eq:prop}
\en
with $u_n$ given by the Hertzian result, Eq. (\ref{eq:hertz}).

There are other types of motions possible for a contact. One is torsional
motion, corresponding to the rotation of the ball about an axis normal to the
plane of the contact. Torsional motion leads to couples that vanish as the
contact size goes to zero, so we will disregard this type of motion. Rotation
of the particle about an axis parallel to the plane of the contact, on the
other hand, leads to a rolling motion, in which the center of the contact moves
parallel to the underlying plane.

Unfortunately, the strains corresponding to rolling motion have not been
successfully computed analytically, even for small Poisson ratio, although a
number of approximate solutions have been proposed. For rolling at fixed normal
force, there is always a region of microslip at the trailing edge of the
contact \cite{johnson}. However, it does not necessarily follow that a contact
that rolls while being proportionally loaded experiences microslip, since the
normal force will be increasing in this case. In any case, proportional loading
will lead to a unique relation between
$u_t/u_n$ and
$T/N$, where $u_t$ is defined as the transverse displacement of the ball center
of mass with respect to the contact, not with respect to its initial position.
For our purposes, it will be sufficient to use Eq. (\ref{eq:prop}) as an
approximate form, since the qualitative properties of the solution are
relatively independent of the exact form of the compliance. 


In order to insure that the contacts are proportionally loaded, we consider the following
experiment. First the ball is placed into the groove at zero gravity so that it barely
touches both walls of the groove. Then the gravitational constant is slowly increased from zero
to its physical value. Since the external force during this loading increases and
does not change direction, we expect $T \propto N$ at each contact with
$N$ monotonically increasing.

Let us suppose that the center of mass of the ball moves a distance $Z$ along a path 
that makes an angle $\beta_L$ with the left-hand wall and $\beta_R$ with the right-hand wall
(Note that $\beta_L + \beta_R = \pi - 2 \gamma$.)
During the loading, the ball may also rotate as a rigid body by an angle
$\phi$, with $\phi > 0$ denoting counterclockwise rolling. Finally, the position of 
the contacts with the wall will in general move by $u_{c;\LR}$ on the left and right-hand
walls respectively; $u_c > 0$ corresponds to contact motion down the wall.

Now we must specify the normal and tangential displacements of the center of
mass of the ball with respect to the contact positions on the two walls. The
normal displacements $u_{n;\LR}$ are relatively simple to write:

\eq
u_{n;\LR} = Z \sin \beta_{\LR}. \label{eq:norm}
\en
The tangential displacements are a little bit more complicated, because of rolling,
and because the left and right-hand contacts may have slipped by $d_{\LR}$
during the loading, respectively. Then, $u_{c;L} = -Z \upsilon +
d_L$ and $u_{c;R} = Z \upsilon + d_R$, where $\upsilon \equiv R \phi/Z$.  
Thus, the tangential displacements $u_{t;\LR}$ are given by

\eq
u_{t;\LR} = Z (\cos \beta_{\LR} \pm \upsilon) - d_{\LR}. \label{eq:tang}
\en

We are now using the five kinematic variables $Z$, $\beta_L$,
$\upsilon$, and $d_{\LR}$. The proportional loading assumption insures
that whether the contacts are loaded with or without rolling, the forces still
obey

\eq
\frac{u_{t;\LR}}{u_{n;\LR}}=\frac{T}{N},
\en
where Amonton's law insures that this is true for sliding contacts as well.


\paragraph*{Rolling solutions}

We first consider solutions in which both contacts roll without sliding: One of 
the contact points retrogrades up its contact plane, while the other rolls down 
its plane. Thus we have
Eqs. (\ref{eq:norm}) for the normal displacements, while Eqs.
(\ref{eq:tang}) become

\eq
u_{t;\LR} = Z (\cos \beta_{\LR} \pm \upsilon), \label{eq:rollt}
\en

In principle, we have a uniquely soluble system, since we have the three
variables $Z$, $\beta_L$ and $\upsilon$ describing the kinematics, with three
conditions of static equilibrium. However, we must also impose Eqs. (\ref{eq:Amont})
constraining the values of $T/N$. Thus there will be some restricted
region of the $\gamma - \delta$ plane in which solution of the Eqs.
(\ref{eq:equil1}-\ref{eq:equil3}) is possible. Substituting Eqs.
(\ref{eq:norm}) and (\ref{eq:rollt})
into Eqs. (\ref{eq:equil1}-\ref{eq:equil3}), we obtain

\eq
\left (\frac{\sin \beta_R}{\sin \beta_L} \right )^{\frac{3}{2}}
 \!\!\!\!=
\frac{\sin \theta_L}{\sin \theta_R} +  \frac{\left( \cos
\beta_L+\upsilon\right)\left(\cos \theta_R -\cos \theta_L\right)}{\sin
\beta_L \sin \theta_R},
\label{eq:tran1}
\en
and
\eq
\left( \frac{\sin \beta_R}{\sin \beta_L} \right )^{\frac{1}{2}}  \!\!\!\!= 
\frac{\cos \beta_L + \upsilon}{\cos
\beta_R - \upsilon}.
\label{eq:tran2}
\en

\begin{figure}
\centerline{\epsfxsize=3.1in \epsffile{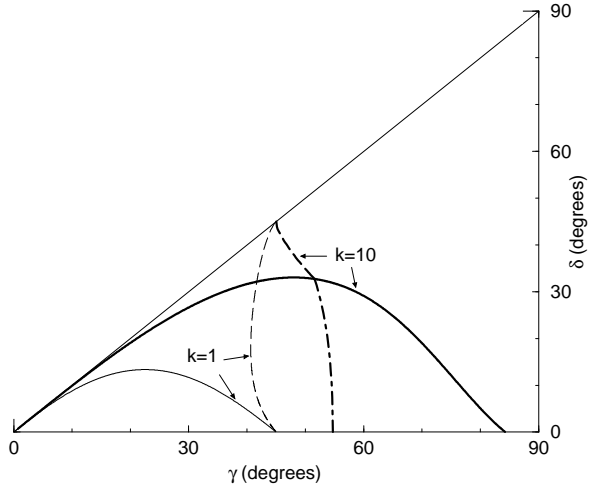}}
\caption{Boundaries between different regimes of behavior for coefficient 
of friction $k=1$ (thin lines) and $k=10$ (thick lines), as a function of 
$\gamma = \frac{1}{2} (\theta_L + \theta_R)$ and $\delta =
\frac{1}{2} (\theta_L - \theta_R)$. Solid lines mark the boundary between 
the elastic double-rolling regime below and the plastic
rolling-sliding regime above. The dashed and dot-dashed lines mark the 
boundaries between clockwise (low-$\gamma$) and counter-clockwise (high-$\gamma$) rolling.}
\label{graph}
\end{figure}

It is straightforward to solve these equations for
$\beta_L$ (and thus $\beta_R$) and for $\upsilon$. For
$\theta_L = \theta_R$ (or $\delta = 0$), $\beta_L = \beta_R = 
\frac{\pi}{2}-\gamma$ and $\upsilon = 0$, i.e., there is no rolling and 
both contacts are stuck. Furthermore,
\eq
\frac{u_{t;L}}{u_{n;L}} = \frac{u_{t;R}}{u_{n;R}} = \cot \beta_L = \tan \gamma
= \frac{T_L}{N_L} = \frac{T_R}{N_R}<k.
\en
Hence, the solution for $\delta=0$ can only exist for $\gamma < 
\arctan(k)$, because for $\gamma > \arctan(k)$, the values of $T/N$ corresponding to 
this solution exceed $k$. Thus, emanating from the point $(\gamma,\delta) = (\arctan(k),0)$ 
is a boundary beyond which double rolling is impossible, because it would require
illegal values of $T/N >k$. This bound is exceeded first on the steeper of the two
walls. This boundary, shown as a solid line in Fig.~\ref{graph}, rejoins the 
$\gamma$-axis at $\gamma =0$, asymptotically merging with the line $\delta = \gamma$.

There is an additional solution for which $\upsilon =0$ (no rolling), with $\beta_L 
\ne \beta_R$, that exists for $\frac{\pi}{4} \le \gamma \le \arctan(\sqrt{2})$. 
This additional solution always has at least one of the ratios $T/N \ge
\sqrt{2}$, so it can only appear for $k > \sqrt{2}$, and constitutes a boundary between a
regime in which the ball rolls clockwise (i.e., down the steeper surface) for smaller $\gamma$,
and a regime in which the ball rolls counter-clockwise (i.e., up the steeper surface) for larger
$\gamma$. The values of $(\gamma,\delta)$ for which this solution occurs is shown for 
$k=10$ as a dot-dashed line in Fig.~\ref{graph}.
 
\paragraph*{Rolling-sliding solutions}

Beyond the boundary at which the steeper wall has $T=kN$, the only solutions possible are 
``rolling-sliding" solutions, in which one contact rolls without sliding while the other 
slides. Assuming that the left-hand wall slides, we have
\eq
\frac{T_L}{N_L} = k.
\en
The equations of static equilibrium then imply
\eq
\frac{T}{N_R} = \frac{\sin \theta_R}{\cos\theta_R - \cos\theta_L + \frac{1}{k}
\sin \theta_L} \equiv \mu,
\en
where $\mu$ depends only upon the geometry and upon the coefficient of friction
$k$. Note that for $\theta_R < \theta_L$, $\mu < k$, while for $\theta_R >
\theta_L$, $\mu > k$. Thus, sliding always occurs on the steeper wall, which 
is indeed the left-hand wall for $\delta \ge 0$. This is the simplest continuation of
the double-rolling solution, which failed at the boundary through incipient
sliding at the steeper wall. We have 
\eq
\frac{N_L}{N_R} = \left ( \frac{\sin \beta_L}{\sin \beta_R} \right
)^{\frac{3}{2}} = \frac{\mu}{k} ,
\en
which can be uniquely solved for $\beta_L$ and $\beta_R$. In addition,
the criterion that the ball rolls at the right-hand contact implies that
\eq
\frac{T}{N_R} = \mu = \frac{\cos \beta_R - \upsilon}{\sin \beta_R},
\en
which allows us to solve for $\upsilon(\beta_R)$. 

The physical constraint $d_L > 0$ sets the boundary on the region with a rolling-sliding
solution, which, for $\gamma < \arctan(k)$, coincides with the upper boundary of
the double rolling  solution discussed in the previous section. For $\delta = 0$
we have
$\upsilon = 0$ and $T/N_R =k$, and this solution collapses onto a double sliding
solution  for $\gamma > \arctan(k)$. In addition, there is a line of solutions 
with $\upsilon =0$, corresponding to one contact sliding and the other being stuck. 
For $k < \sqrt{2}$ this line emanates from
$(\gamma,\delta) = (\arctan(k), 0)$ and ends at $(\gamma,\delta) =
(\frac{\pi}{4},\frac{\pi}{4})$ . For $k > \sqrt{2}$, this line emanates from the point on
the boundary of the double rolling region for which $\upsilon=0$, and ends at $(\gamma,\delta) =
(\frac{\pi}{4},\frac{\pi}{4})$ (dashed lines in Fig.~\ref{graph}). As for the double rolling
$\upsilon=0$ line, this line separates the low-$\gamma$ clockwise rolling from the
high-$\gamma$ counterclockwise rolling.


Much discussion of the statics of granular media has
emphasized the two limits of plastic and linear elastic behavior. 
Clearly the rolling-sliding regime of the ball in the groove is a good analog
to plastic behavior--in this regime, the problem is {\it isostatic}, meaning
that the number of independent forces is identical to the number of
constraints. In the double rolling regime, the state of the ball is specified by its
displacements, as in conventional elasticity theory. However, minimization of the
strain energies corresponding to the contact forces plays no role in
resolving the stress indeterminacy.

It is instructive to consider the relationship between the forces and the
underlying degrees of freedom $Z$, $\beta_L$, $\upsilon$,  and $d_{\LR}$. The four forces
$T_{\LR}$ and $N_{\LR}$ are functions of these five underlying variables. In the double-rolling
regime, $d_{\LR}= 0$, and there are only three active variables, which are then uniquely fixed
by the equations of static equilibrium. We call this an isokinetic configuration, because the
number of kinetic degrees of freedom equals the number of constraints enforcing equilibrium. 

On the other hand, as one passes into the isostatic rolling-sliding regime, one
acquires a new degree of freedom $d_L$, so that there are now four underlying
active degrees of freedom. However, one of these degrees of freedom is used to
fix $T_L = k N_L$, so in practice there are three degrees of freedom remaining
to satisfy the equations of static equilibrium.

We are grateful to P.M. Chaikin and D. Levine for stimulating discussions.

\end{document}